\documentclass[12pt,a4paper]{conference}

\usepackage{fancyhdr}
\usepackage{graphicx,amsmath,amssymb,cite}
\usepackage{multind}
\newcommand{\beq}{\begin{equation}}
\newcommand{\eeq}[1]{\label{#1}\end{equation}}
\newcommand{\eeqn}{\end{equation}}

\newcommand{\beqa}{\begin{eqnarray}}
\newcommand{\eeqa}[1]{\label{#1}\end{eqnarray}}
\newcommand{\eeqan}{\end{eqnarray}}

\let\bar=\overbar

\newcommand{\Dslash}{\not{\hbox{\kern-4pt $D$}}}
\newcommand{\dslash}{\not{\hbox{\kern-2pt $\del$}}}

\newcommand{\ee}{e^+e^-}

\newcommand{\msb}{{\bar{\ssstyle M \kern -1pt S}}}

\newcommand{\be}{\begin{equation}}
\renewcommand{\ee}{\end{equation}}
\newcommand{\ba}{\begin{eqnarray}}
\newcommand{\ea}{\end{eqnarray}}

\begin{document}

\Chapter{Illuminating the $1/x$ moment of \\ parton distribution 
functions}{Illuminating $\int dx \frac{f(x)}{x}$}{Brodsky, 
Llanes-Estrada and Szczepaniak}
\vspace{-6 cm}\includegraphics[width=6 cm]{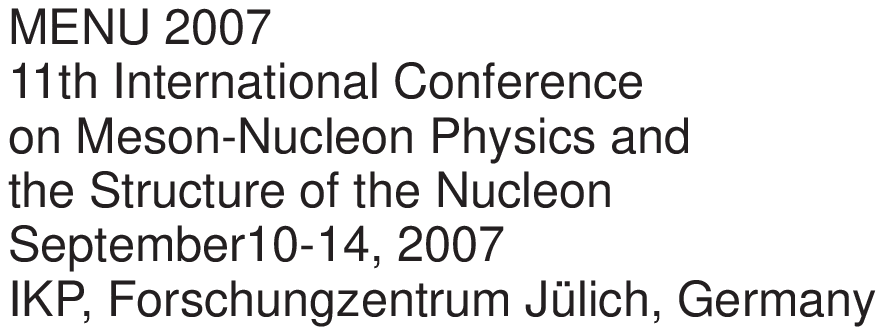}
\bigskip\bigskip
\vspace{4 cm}

\addcontentsline{toc}{chapter}{{\it N. Author}} \label{authorStart}

\begin{raggedright}


Stanley J. Brodsky $^{\star}$, \underline{Felipe J. 
Llanes-Estrada}$^{\%}$ and Adam P. Szczepaniak$^{\#}$ 
\bigskip\bigskip


$^{\star}$ 
Theory Group, Stanford Linear Accelerator Center, 2575
Sand Hill Road, 94025 Menlo Park, California, USA. {\tt 
sjbth@slac.stanford.edu} \\
$^{\%}$ Depto. F\'{\i}sica Te\'orica I, Fac. Cc. F\'{\i}sicas,
Universidad Complutense de Madrid, 28040 Madrid, Spain. 
\footnote{Talk and poster presented at MeNu07, part of a
work in preparation.}. {\tt fllanes@fis.ucm.es}\\
$^{\#}$ Nuclear Theory Center, Indiana University, 2401 Milo Sampson
Lane, Bloomington, IN 47408 Indiana, USA. {\tt aszczepa@indiana.edu}

\end{raggedright}

\begin{center}
\textbf{Abstract}
\end{center}
The  Weisberger relation, an exact statement of the parton model, 
elegantly relates a high-energy physics observable, the $1/x$ moment of 
parton distribution functions, to a nonperturbative low-energy 
observable: the dependence of the nucleon mass 
on the value of the quark mass or its corresponding quark condensate. 
We show that contemporary fits to nucleon structure functions fail to
determine this $1/x$ moment;  however,  deeply virtual Compton 
scattering can be described in terms of  a novel  $F_{1/x}(t)$ 
form factor which illuminates this physics.
An analysis of exclusive photon-induced processes in terms of the parton-nucleon scattering
amplitude with Regge behavior  reveals a failure of the
high  $Q^2 $  factorization of exclusive
processes at low $t$  in terms of  the Generalized Parton-Distribution Functions
which has been widely believed to hold in the past.
We emphasize the need for more data for the DVCS process at large  $t$   in future or
upgraded facilities.

\section{The Weisberger relation}

The importance of the $1/x$ moment of parton distribution functions
(pdf's) was stressed in a 1972 paper by W. Weisberger
\cite{Weisberger:1972hk}.  There he derived a relation between
the $1/x$ moment and the derivative of the squared proton mass with respect
to the squared parton mass defined at the
same renormalization scale $\mu$. In modern notation and normalization
\cite{Martin:1998sq}, Weisberger's result reads
\begin{equation} \label{Weisberger}
\frac{\delta M_N^2}{\delta m_i^2(\mu)} = \int_0^1 \frac{dx}{x}
 \left( f_i(x)_\mu + \bar{f}_i(x)_\mu \right) \ .
\end{equation}
Here $f_i$ is the $i$th-quark distribution function, and since CPT
invariance implies that the mass of quark and antiquark are equal and
must be varied together, we have also included the antiquark pdf $\bar{f}$.

With the advent of QCD and the Hellmann-Feynman theorem, one can see 
that
Weisberger's result holds simply by noting that in light front
quantization the Hamiltonian  contains a kinetic energy term
\begin{equation}
M^2_{\rm kin}=\sum_i {k^2_\perp + m^2_i\over x_i}.
\end{equation}
where $x =k^+/P^+ = (k^0+ k^z)/(P^0 + P^z)$ is the light-front momentum 
fraction.
After regularization and renormalization\cite{Paston:2000fq},
a $c_8$ mass counterterm appears, but no mass dependence in any of the
other counterterms, so that the formal manipulation of the
Hellmann-Feynman theorem remains valid in the regulated Hamiltonian.

Upon taking the expectation value $\langle \delta M^2/ \delta m_i^2
\rangle_{\psi}$, the trivial integration over
the $k_\perp$ transverse variables leads to the $\int f/x$ result.   
Thus the Weisberger relation Eq. \ref{Weisberger} relates the 
variation of the proton mass to the quark mass terms which appear 
specifically in the LF kinetic energy. (The Weisberger relation is an 
exact statement of the parton model, but in full QCD there could be an 
additional term due to an implicit mass dependence of the fields, which 
is under investigation).

In chiral perturbation theory the quark mass dependence of the nucleon
mass is parameterized in terms of a contact term with an unknown constant
$c_1$ \cite{Kubis:2007iy}
\be
M_N(m_q)=M_N(0)-4c_1m_\pi^2 - \frac{3g_A^2 m_\pi^3}{32\pi f_\pi^2}+
O(m_\pi^4) \ .
\ee
The constant counterterm $4c_1m_\pi^2$ is related to the $\sigma$ 
term of the 
nucleon, the expectation value of a scalar current.
The accurate determination of the ${1/x}$ moment of pdf's,  including its scale
dependence, would thus allow a determination of the $\sigma$ term 
independently of chiral perturbation theory, or alternatively, provide a 
constraint between high and low-energy physics which tests the way  
mass terms enter the QCD Hamiltonian. 
One can also combine independent 
evaluations of the $\sigma= \hat{m} \partial m_N / \partial\hat{m}$ term 
and the $1/x$ moment to provide an evaluation of the
quark mass via (for isospin averaged light quarks)
\begin{equation}
\hat{m}^2=  \frac{M_N \sigma}{\langle 1/x \rangle}\ .
\end{equation}

The spontaneous chiral symmetry breaking pattern of QCD also allows us to 
write down a new sum rule for the pion distribution function by combining the Weisberger relation 
with the {Gell-Mann-Oakes-Renner relation}, which links 
the quark mass to 
the quark condensate in the chiral limit:
\begin{equation}
\int_0^1 \frac{dx}{x} f^\pi_u(x) =  \frac{\langle \bar{\Psi} \Psi
\rangle_{m=0}^2}{m_\pi^2 f_\pi^4} \ .
\end{equation}
The left and right-hand sides vary with the 
scale in the same way (since $m\langle\bar{\psi}\psi\rangle$ is 
renormalization-group invariant). This result is independent of 
the light-front formalism  since $\delta M_N^2/\delta m_q^2$ 
can be studied in any framework. The light-front formalism,  however, 
provides the tools needed to demonstrate the relation. This 
new sum rule can be of use to constrain models of the pion as well as 
Deep Inelastic Scattering data.

\section{Regularization of the Weisberger relation}
The parton distribution functions measured in deep inelastic lepton 
scattering are observed to diverge at small
$x_{bj}$ due to the
Regge behavior of the forward virtual Compton amplitude and simple analytic
arguments.
In fact, modern fits to deep inelastic scattering data at small $Q^2$ routinely employ
a parameterization of pdf's which is a simple variation of
the Kuti-Weisskopf model  \cite{Kuti:1971ph}, namely
\begin{equation} \label{KWfits}
xf_i(x) = A_i x^{\eta_i} (1-x)^{\lambda_i}(1+\gamma_i \sqrt{x}
+ \epsilon_i x)
\end{equation}
where all parameters are left free for the fit. 
The phenomenology of deep inelastic
scattering generally requires $\eta$ to be smaller than 1 for several pdf's.
In fact,
for the valence flavors, $\eta_i=1-\alpha(0)$, a typical Regge intercept
$\alpha(0)=1/2$ makes the integral in eq. \ref{Weisberger} to be
manifestly divergent. This is the case for the GRV98 pdf set
\cite{Gluck:1998xa}
which has exponents $-0.85$ and $-0.52$ for the light sea and valence
pdf's respectively. Notice that the $\sqrt{x}$ in eq. \ref{KWfits} gives
rise to subleading Regge power laws.
For the MRST98  \cite{Martin:1998sq}  pdf sets, an also widely used
alternative, the power-law exponents have higher variation around
classical Regge theory and the $u$ proton's valence component has a
somewhat high intercept $\alpha_u(0)\in(0.53,0.59)$, the $d$ valence
component being definitely at odds with other phenomenology with
$\alpha_d(0)\simeq 0.73$ as large as the sea component. The subleading
Regge behavior is also  given by the $\sqrt{x}$ factor in eq.
\ref{KWfits}, and having an  intercept larger than zero, it also causes
a divergence. In both GRV98, MRST98 sets the gluon pdf behaves as a
valence-like parton with a very small intercept at this low scale,
indication of the gluon degrees of freedom being truncated at low energy
\cite{Alkofer:2006xz}.

The  Weisberger relation is thus formally divergent and needs to 
be properly regulated. This can be done either by analytical 
continuation from large $t$ where the amplitudes are convergent  
\cite{Damashek:1969xj} or by studying the spectral representation of the 
parton-nucleon scattering amplitude which underlie the parton-distribution 
functions. Both topics will be discussed briefly below, but 
meanwhile let us give the correctly regulated relation as found by 
Brodsky, Close, and Gunion(BCG)   \cite{Brodsky:1973hm},
\begin{eqnarray} \label{Weisbergerregulated}
\frac{\delta M_N^2}{\delta m_i^2} = \int_0^1 \frac{dx}{x}
 \left( f_i - f_i^{\rm Regge} + \bar{f}_i-\bar{f}_i^{\rm Regge} 
\right)(x)
\ - \sum_\alpha \frac{\gamma_\alpha}{\alpha(0)}
- \sum_{\bar{\alpha}} 
\frac{\bar{\gamma}_{\bar{\alpha}}}{\bar{\alpha}(0)}
\\ \nonumber
f_i^{\rm Regge}(x) = \sum_\alpha \gamma_\alpha x^{-\alpha(0)}\ \ 
\alpha(0)>0
\\ \nonumber
\bar{f}_i^{\rm Regge}(x) = \sum_{\bar{\alpha}} 
\bar{\gamma}_{\bar{\alpha}}
x^{-\bar{\alpha}(0)}\ \ \bar{\alpha}(0)>0 \ .
\end{eqnarray}

Notice that the particular form of this subtraction entails that 
there can be no Regge pole with exactly $\alpha=0$ in the input 
fit (this is set by the BCG choice of the subtraction 
point in the parton-nucleon scattering matrix formalism).  As a 
consequence we cannot currently examine the pion sum rule with standard 
pion distribution functions \cite{Sutton:1991ay} since $\alpha=0$ 
constant terms do appear in those parametrizations. It may be possible 
to develop an equivalent formula with a different  
subtraction point to avoid this inconvenience. 

\begin{table}
\caption{Weisberger integral $\int_0^1 dx \frac{f(x)}{x}$ for MRST98
\cite{Martin:1998sq} and GRV parton distribution functions. Following
BCG, we have analytically continued in $t$ as in 
eq.
(\ref{Weisbergerregulated}) by adding and subtracting the divergent
Regge terms. The sets are taken at low-energy input  scales  $1 
GeV^2$ (MRST) and $0.26(0.4) GeV^2$ for the LO(NLO) GRV set. The latter 
has no strange sea component at this low scale.
\label{tableweisberger}}
\begin{tabular}{|c|ccc|cc|} \hline
quark & MRST &MRST &MRST &
LO &NLO \\
flavor & Low gluon & Central gluon &  Upper gluon & GRV & GRV
\\
\hline
$u$       & 34  & 8.6 & 12   & 133 & 26\\
$\bar{u}$ &-1.3 &-5.2 & -7.1 & 62  & 5.8\\
$\frac{\delta M_N^2}{\delta m_u^2}$ & 33 & 3.4 & 4.9 & 195 & 32 \\
\hline
$d$       & 0.98&-0.4 & 0.33 & -20 & -5.7\\
$\bar{d}$ &-0.46&-0.75&-1.8  & -62  & -17 \\
$\frac{\delta M_N^2}{\delta m_d^2}$ & 0.52 & -1.1 & -1.5 & -82 & -22\\
\hline
$s$       & -0.43 & -1.5 & -2.2 & 0 & 0\\
$\frac{\delta M_N^2}{\delta m_s^2}$ & -0.86 & -3.0 & -4.4 & 0 & 0\\
\hline
$g$       & $\simeq 600$ & $\simeq 400$ & $\simeq 2900$ & 10 & 12\\
\hline
\end{tabular}
\end{table}

The result of computing the properly regularized $1/x$ moment from Eq. \ref{Weisbergerregulated}
 for a few
standard pdf sets is given in table \ref{tableweisberger}. As can be
seen, there is considerable spread in the results, and much room for
improvement in  the determination of the moments.

\section{The $1/x$ Form Factor of the Nucleon}
An important empirical way to access the $1/x$ moment of parton distribution 
functions is by utilizing the forward limit 
of the generalized parton distribution (GPD) functions measured in deeply virtual Compton scattering (DVCS); specifically,
\be
H(x,\zeta=0,t=0)=f(x)
\ee
so that the  
{$F_{1/x}(t)$ form factor} defined by
\be
F_{1/x}(t) = \sum e_q^2 \int^1_0 dx \frac{H(x,0,t)}{x}
\ee
should take in the $t\to 0$ limit a value given by a sum of 
the $1/x$ moments for various flavors. Unfortunately this equation is 
known to be rigorously valid for sizeable $t$ only.  In that case, the 
$F_{1/x}(t) $ form factor is accessible via DVCS; 
the required DVCS amplitude in the GPD-collinear factorization formalism is given by
\begin{equation}\label{amplitude}
{\mathcal{M}}^{++}(s,t,Q^2)= \frac{-e_q^2}{2}
\frac{\sqrt{1-\zeta}}{1-\zeta/2}
\int_{\zeta-1}^{1} dx\left[ \frac{1}{x-i\epsilon} +
\frac{1}{x-\zeta+i\epsilon}  \right]  H(x,\zeta,t) \ .
\end{equation}
(Here we have ignored the contribution of the $E(x,\zeta,t)$ GPD, 
and work in the asymmetric frame).

The experimental determination of the $F_{1/x}(t) $ form factor would in principle allow 
an analytic continuation in $t$ to $t=0$, thus providing
$1/x$ moment. However, as noted in the next section, it is not trivial 
to carry out such an extrapolation through the low $t$ region due to Regge 
divergences.

\begin{figure}
\includegraphics[width=10cm,angle=-90]{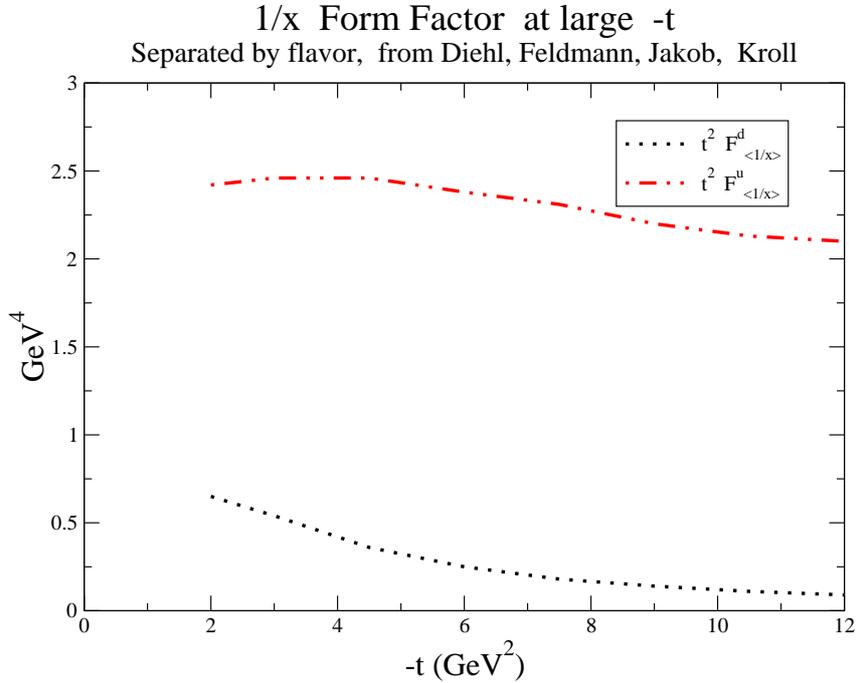}
\caption{\label{1overxDiehl}
An evaluation of the $1/x$ form factor of the proton assuming the Gaussian light-front constituent quark model utilizing
flavor-separated form factors obtained from a set of GPD's fit to a
number of conventional Dirac and Pauli form factors. Data from
\cite{Diehl:2004cx} is given only at large $-t$ by direct calculation.}
\end{figure}

A prediction for the $1/x$ form factor of the nucleon has been given 
for  a particular model, the Gaussian light-front constituent quark model;  this is illustrated in figure 
\ref{1overxDiehl}.

\section{Loss of collinear factorization in deeply virtual Compton scattering}

It has been recently shown\cite{Szczepaniak:2006is,Szczepaniak:2007af}
that the DVCS amplitudes can be most efficiently described 
in terms of $t$-channel Regge exchanges.   The analysis proceeds along the lines of 
ref. \cite{Brodsky:1973hm} by employing a representation of the leading twist  amplitude as an integral over the underlying parton-nucleon scattering 
amplitude  \cite{Landshoff:1970ff}. The DVCS amplitude can then be written in terms of a subtracted spectral 
representation
\be
A^{\pm}=\sum_n c_n (2\pi)^4 \int dm^2 \left[
\frac{\rho_n}{s_{pp}-m^2+i\epsilon}
-\frac{\rho_R^n}{-m^2+i\epsilon} \pm (s_{pp}\to u_{pp})\right]
f(k^2,k^{'2})
\ee
The convergence  needed for the handbag diagram is
provided by the regularization procedure 
$I_n= (m^2)^n \frac{d^n}{(d m^2)^n}$, $n>1$
and the functional form of $f$.
The Regge behavior follows from  the form $\rho^R\propto 
(m^2)^{\alpha(t)}$ (under the assumption that the quark-nucleon 
matrix element has standard hadronic physics properties  \cite{Landshoff:1970ff}).

\begin{figure}
\includegraphics[width=10cm,angle=-90]{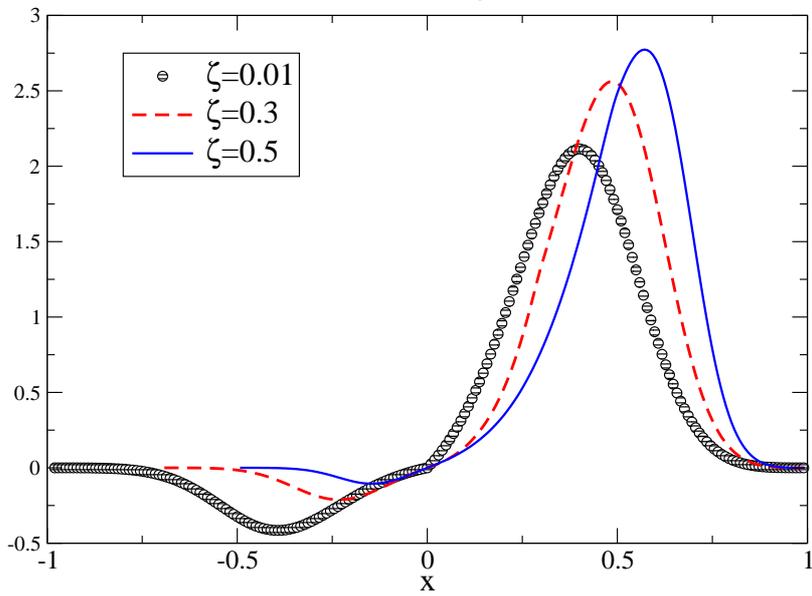}
\caption{\label{fig:gpddiquark}
The GPD $H(x,\zeta,t)$ at fixed $-t=1 GeV^2$ in a perturbative 
quark-diquark model  \cite{Brodsky:2005vt}
with masses 400 $MeV$ and 800 $MeV$ for the quark 
and diquark, a vertex coupling $g=25$ and both $u$ and $s$ channel 
exchanges (the latter, covering the antiquark region $x<0$, have been 
suppressed by an ad-hoc 0.2 factor since this is a valence-like model)}.
 \end{figure}

At large $t$, one can eschew Regge behavior and think of such 
representation as a dispersive integral over diquark exchanges of 
varying mass. The form of the resulting nucleon GPDs (see figure \ref{fig:gpddiquark})
are similar to those found in ref. \cite{Tiburzi} and ref. \cite{Ji} for the GPD of the pion.

However, if one now proceeds to study the low-$t$ region, one finds 
Regge poles at the break-points of the GPD, for example, approaching 
$x=\zeta$ from higher values of $x$ one finds
\be
H(x\to\zeta^+, \zeta, t\simeq 0) \to \left[
 \pi^2 m_q^2 I_{n-1}\beta \int_0^\infty
\frac{d\phi \phi^\alpha}{(\phi^2+ m_q^2)^2}\right]\
\left[
\frac{(1-\zeta)^\alpha}{1+\alpha}(x-\zeta)^{-\alpha}-
\frac{\zeta^{-\alpha}}{1+\alpha}
\right]
\ee
where the function on the left bracket will be discussed in detail in our upcoming 
publication. The function on the right however shows clearly the Regge 
pole $(x-\zeta)^{-\alpha}$. As a result, whereas the DVCS amplitude correctly exhibits Regge scaling 
in $s$, its $Q^2$ dependence does not track with the same power; 
the amplitude at nonzero-$t$ thus cannot scale with Bjorken $\zeta$ alone.  
Such Regge contributions were not contemplated in the original proof of the collinear
factorization theorem   \cite{Collins:1998be} and thus apparently make it fail. 
Current models, such as the one presented in 
in figure \ref{fig:gpddiquark}, which have soft behavior at the 
break-points also must be improved.

\section{The $J=0$ fixed-pole in Compton Scattering.}

In  
{Regge} theory, hadron scattering amplitudes scale 
as $s^\alpha(t)$, 
where the exponent of the Regge pole evolves with $t$.  A fixed pole at 
$J=\alpha=0$ corresponds to a constant real amplitude. 
Such behavior was 
proposed  \cite{Creutz:1968ds} and  found 
  \cite{Damashek:1969xj} in Compton scattering in the late sixties. In their analysis
Damashek and Gilman\cite{Damashek:1969xj}  used the forward dispersion relation for the Compton amplitude and measurements of total photoabsorption cross section $\sigma(\gamma p \to X)$  to show that the forward Compton scattering on the proton has a  $J=0$ contribution.
A formal proof that the Compton amplitude must present
fixed pole behavior was given in ref.   \cite{Zee:1972nq}.  Physically it arises from the local four-point seagull interaction in scalar QCD or  from the instantaneous fermion exchange interaction 
in the light-front QCD Hamiltonian  \cite{Brodsky:1971zh,Brodsky:1997de}.  The $J=0$ contribution to the DVCS amplitude is thus independent of $s$ for any photon virtuality and any momentum transfer $t$.

\begin{figure}[h]
\includegraphics[width=6cm]{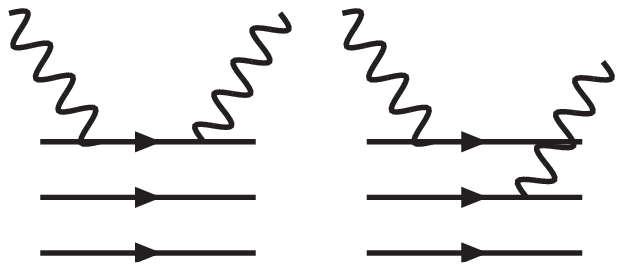}
\caption{In the handbag mechanism the left diagram dominates the
cross-section over all others such as, say, the right one. This is 
testable through the ratio of the Compton scattering cross sections of 
the neutron over the proton. }
\label{fig:whichquarkhit}
\end{figure}

In general, the contribution to Compton scattering (real or virtual)  which is directly sensitive to 
the $1/x$ moment can be identified with the  ``handbag" diagram in QCD where the 
incoming and outgoing photons interact on the same valence quark line.  
Note that 
in the case of fixed $\theta_{CM}$ angle Compton scattering,
where $t, u,$ and $s$ are all large,  the outgoing photon can be equally well emitted 
by another valence quark (see figure \ref{fig:whichquarkhit}).
Therefore, Compton scattering at fixed angle does not isolate the handbag diagram.
The optimum experimental approach is thus to work in the Regge regime 
for DVCS.
As shown in ref.   \cite{Brodsky:1973hm}, the 
$J=0$  {fixed pole} has $t$ 
dependence given precisely by the $1/x$ form factor.

Thus, a good experimental strategy to extract the $J=0$ $F_{1/x}(t)$ form factor is to fix 
$t$ and let $s$ increase until only the constant fixed-pole amplitude 
remains. Since the contribution to the DVCS amplitude is real, it can be extracted from  
interference with the Bethe-Heitler amplitude  \cite{Brodsky:1971zh}. 
In addition, if one 
wants to interpret this form factor in terms of a moment of GPD's, one 
needs to demand $Q^2>>-t$. 
An upgraded Jefferson Laboratory with a 12 $GeV$ beam should be able to reach
perhaps $s\simeq 40 \ GeV^2$, $Q^2 \simeq 6 ~GeV^2$, $t\simeq -3\ GeV^2$
and thus
should be able to report a first measurement in a regime where
the virtual Compton amplitude should become $Q^2$ and $s$ independent.
The extracted $t$ dependence would provide the first measurement of the
$F_{1/x}(t)$  nucleon form factor. It is also possible that current measurements by the Hall A collaboration at JLab  \cite{Danagoulian:2007gs} of 
$R_V(t)$ in real Compton scattering also yield the same physics, but 
there is no kinematic limit where one can perform the needed checks.
The kinematically stringent Regge limit of DVCS at sizeable $t$ provides 
further motivation for a future electron-proton collider.

\section*{Acknowledgments}
We are indebted to many colleagues for valuable discussions, among them 
Tim Londergan.
Work supported by grants FPA 2004-02602, 2005-02327,
PR27/05-13955-BSCH (Spain),  DOE contracts DE-FG0287ER40365 
DE-AC02-76SF00515, and NSF under grant PHY-0555232 (USA).

\appendix


\end{document}